\documentclass[3p,authoryear]{elsarticle}

\usepackage{natbib}

\usepackage{graphicx}

\usepackage{amsmath}

\begin{document}

\begin{frontmatter}



\title{Heterogeneity of cells may explain allometric scaling of metabolic rate}


\author[KIT]{Kazuhiro Takemoto\corref{cor}}
\ead{takemoto@bio.kyutech.ac.jp}

\address[KIT]{Department of Bioscience and Bioinformatics, Kyushu Institute of Technology, Iizuka, Fukuoka 820-8502, Japan}
\cortext[cor]{Corresponding author}

\begin{abstract}
The origin of allometric scaling of metabolic rate is a long-standing question in biology. Several models have been proposed for explaining the origin; however, they have advantages and disadvantages. In particular, previous models only demonstrate either two important observations for the allometric scaling: the variability of scaling exponents and predominance of 3/4-power law. Thus, these models have a dispute over their validity. In this study, we propose a simple geometry model, and show that a hypothesis that total surface area of cells determines metabolic rate can reproduce these two observations by combining two concepts: the impact of cell sizes on metabolic rate and fractal-like (hierarchical) organization. The proposed model both theoretically and numerically demonstrates the approximately 3/4-power law although several different biological strategies are considered. The model validity is confirmed using empirical data. Furthermore, the model suggests the importance of heterogeneity of cell size for the emergence of the allometric scaling. The proposed model provides intuitive and unique insights into the origin of allometric scaling laws in biology, despite several limitations of the model.
\end{abstract}

\begin{keyword}
Allometric scaling \sep Metabolic rate \sep Geometry model \sep Fractal-like organization \sep Log-normal distribution
\end{keyword}

\end{frontmatter}

\section{Introduction}
\label{sec:introduction}
Metabolic processes are essential for physiological functions and responsible for maintaining life \citep{Takemoto2012,Takemoto2012b}.
The relationship between metabolic rate $B$ and body mass $M$ is an important and interesting topic of scientific inquiry not only for researchers in the field of basic biology but also for investigators in ecology and medical research, and it is well known to approximately obey a power law \citep{West2002,Brown2004}: $B\propto M^{\gamma}$. 
This allometry is positioned as a significant equation in both biology and ecology because it is useful for understanding and for estimating the energy metabolism, lifespan \citep{Spearkman2005}, and animal space use \citep{Jetz2004}; in particular, determination of the scaling exponent $\gamma$ is a long-standing question. 

A pioneering study includes the surface law found by Rubner in the 1880s \citep{White2003}; in particular, Rubner reported that metabolic rate is proportional to $M^{2/3}$ in mammals.
The Rubner's surface law is immediately derived when assuming simple geometric and physical principles: metabolic rate (e.g., heat production rate) is proportional to the rate of energy (e.g., heat) dissipated through body surface because of homeostasis.

Contrary to the prediction from the surface law, in 1932, Kleiber proposed that metabolic rate is proportional to $M^{3/4}$ \citep{White2003,West2002,Spearkman2005,Brown2004}.
Further studies have confirmed that the Kleiber's (i.e., 3/4-power) law is predominant at least in plants and animals (reviewed in \citep{Savage2004}): this allometric scaling is observed in wide-ranging organisms (i.e., microorganisms to elephants). 

\citet{West1997} have proposed a model (West--Brown--Enquist (WBE) model) for explaining the origin of the 3/4-power law.
This model assumes that oxygen and nutrients are transported through space-filling fractal networks of branching tubes, in which the number of capillaries (leaves in the case of plants) is proportional to metabolic rate.
Since the WBE model clearly illustrates the Kleiber's law, it is frequently used for understanding the allometric scaling.

In addition to this, further studies have proposed several alternative models based on the fourth dimension of life \citep{West1999}, optimal transport networks \citep{Banavar1999,Banavar2010,Banavar2014}, and quantum metabolism \citep{Demetrius2010}.

However, these models have several limitations (e.g., \citet{Price2012} have carefully evaluated the WBE model using empirical data).
In particular, \citet{Kozlowski2004} have questioned the universality of the 3/4-power law.
In fact, several studies suggest that the scaling exponent $\gamma$ is variable.
For example, \citet{White2003} have reported that metabolic rate is proportional to $M^{2/3}$ in mammals when considering body temperature, digestive state, and phylogeny.
\citet{Darveau2002} have illustrated that the approximately 3/4-power law results from the sum of the influences of multiple contributors to metabolism and control (i.e., the sum of allometric relationships observed in a number of biological processes).
\citet{Reich2006} found that a linear relationship between the rate of respiratory metabolism and body mass (i.e., $B\propto M$) is observed in plants although \citet{Enquist2007} have refuted this conclusion.
Similarly, the observed similar mean mass-specific metabolic rates across life's major domains \citep{Makarieva2008} also implies that $B\propto M$.
To explain the variability of the scaling exponent, however, the WBE model can be modified \citep{Price2007,Kolokotrones2010}.

Of particular interest is the fact that the scaling exponent varies according to cell size \citep{Kozlowski2003,Starostova2009}.
Thus, \citet{Kozlowski2003} have proposed a simple model (Koz{\l}owski--Konarzewski--Gawelczyk (KKG) model) based on cell size, inspired by the fact that a large part of standard metabolic costs are spent preserving ionic gradients on cell membranes \citep{Szarski1983,Porter1993}.
This model considers an extension of the Rubner's surface law: a hypothesis that metabolic rates are determined by total surface area of cells rather than body surface.
The KKG model can explain the variability of the scaling exponent; especially, the exponent can fall within the range between 2/3 and 1.
For example, $B\propto M^{2/3}$ when cell size is proportional to body size.
On the other hand, $B\propto M$ when cell size is independent from body size.
The variability of the scaling exponent has already well known in terms of geometry \citep{Okie2013,Hirst2014}.

\citet{Brown2005} have argued against the KKG model.
Since cell size is almost independent from body size, as explained by \citet{Kozlowski2003}, the KKG model suggests $B\propto M$, indicating that it contradicts the predominance of the 3/4-power law \citep{Savage2004}.

Is the KKG model or a hypothesis that the total surface area determines metabolic rate not really useful for understanding the allometric scaling law?
In this study, we propose a simple geometry model, an extended KKG model, and show approximately 3/4-power law can be also emerged from this hypothesis by considering the concept of fractal-like (hierarchical) organization of the WBE model.
This result suggests that the hypothesis is evidently useful for understanding allometric scaling of metabolic rate.
Moreover, our model explains both validity of the scaling exponent and ubiquity of the 3/4-power laws, and it suggests that the allometric scaling of metabolic rate is affects by cell size distributions, rather than cell sizes. 

\section{A geometry model}
\label{sec:model}
The KKG model assumes that organisms consist of uniform distributed isometric cells; however, such an assumption may be unsatisfied because organisms show fractal-like (hierarchical) organization, as pointed out by \citet{West1997}.
Furthermore, heterogeneous distributions such as log-normal distribution are widely observed in real-world systems \citep{Limpert2001}.
Multiplicative effects and hierarchical organizations are known to generate log-normal distributions \citep{Kobayashi2011}.
In particular, a simple model of fracture \citep{Yamamoto2013}, originally proposed by A. N. Kolmogorov in 1941, is an instructive example, and it considers hierarchical divisions of one rod.
Thus, we expected that such heterogeneous distributions of cell sizes resulting from the hierarchical organization lead to $B\propto M^{\gamma}$ with $\gamma < 1$ even if (mean) size of cell sizes is almost independent from body size.

\begin{figure}[tbhp]
\begin{center}
	\includegraphics{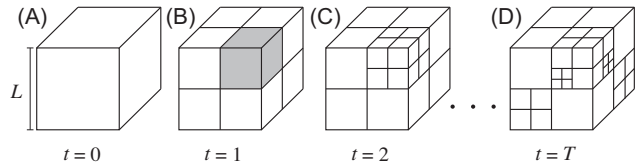}  
	\caption{Schematic diagram depicting the fractal-like cube division (FCD) model.
	(A) Cube with the length of $L$.
	(B) The cube of (A) is divided into 8 cubes (cuboids) with the length of $L/2$. A cube (cuboid) selected at random (i.e., with the probability of $1/8$) is indicated by gray.
	(C) As in (B), the selected cube (cuboid) is separated into 8 cubes (cuboids).
	(D) A state of the cube of (A) after $T$ divisions. 
	}
	\label{fig:cube_model}
\end{center}
\end{figure}

Inspired by this fracture model, we propose a geometry model, called {\it fractal-like cube division (FCD) model} (Fig. \ref{fig:cube_model}).
The FCD model can be interpreted as a 3-dimensional version of the fracture model; however, note that division processes are slightly different between the fracture model and FCD model because of the consideration of fractal-like (hierarchical) organization of cell sizes (e.g., tree branches are more frequently divided than the tree trunk is).

\begin{figure}[tbhp]
\begin{center}
	\includegraphics{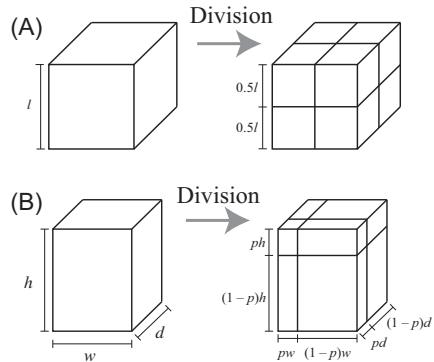}  
	\caption{Division rules of the simple FCD model (A) and general FCD model (B).
	}
	\label{fig:division_rule}
\end{center}
\end{figure}

We first explain a simple case of the FCD model.
In the simple FCD model, the cube with the length of $L$ (Fig. \ref{fig:cube_model}A) is divided according the following procedures.
(i) The cube is divided into 8 cubes (Figs. \ref{fig:cube_model}B and \ref{fig:division_rule}A).
(ii) A randomly selected cube is further separated into 8 cubes (Fig. \ref{fig:cube_model}C).
(iii) The procedure (ii) is repeated until $t=T$ (Fig. \ref{fig:cube_model}D).

However, this division rule of the simple FCD model may be by-the-numbers.
To consider more flexible divisions, we next propose a general FCD model.
This model considers a division of the cuboid $C_{h,w,d}$ composed of the height of $h$, width of $w$, and depth of $d$, in the procedure (ii).
In particular, this cuboid is divided into 8 cuboids according to the parameter $p$ drawn from a probability distribution with the range from 0 to 1 (e.g., uniform distribution $U(0,1)$) (Fig. \ref{fig:division_rule}B):
$C_{ph,pw,pd}$,
$C_{ph,pw,(1-p)d}$,
$C_{ph,(1-p)w,pd}$,
$C_{ph,(1-p)w,(1-p)d}$,
$C_{(1-p)h,pw,pd}$,
$C_{(1-p)h,pw,(1-p)d}$,
$C_{(1-p)h,(1-p)w,pd}$, and
$C_{(1-p)h,(1-p)w,(1-p)d}$.

The general FCD model is equivalent to the simple FCD model when $p=0.5$ at all times.

\section{Results and discussion}
\label{sec:results}
\subsection{Analytical solutions of geometric parameters}
\label{sec:analysis}
We here provide analytical solutions of geometric properties of the simple FCD model; however, the analytical results may be applicable to the general FCD model (see Sec. \ref{sec:evaluation}).

To obtain an analytical solution of the surface area of the simple FCD model using a mean field approximation, we consider the time evolution of the number $n_D(t)$ of cubes with the length of $(1/2)^DL$ when $D>0$ (i.e., $t>0$), where $D$ denotes the cube division number.

$n_D(t)$ increases by 8 when a cube with the length of $(1/2)^{D-1}L$ is selected with the probability of $n_{D-1}(t)/N(t)$, and it decreases by 1 with the probability of $n_D(t)/N(t)$, where $N(t)$ denotes the total number of cubes at time $t$.
Thus, $n_D(t)$ is described as:  
\begin{equation}
\frac{\mathrm{d}}{\mathrm{d} t}n_D(t)=8\frac{n_{D-1}(t)}{N(t)}-\frac{n_D(t)}{N(t)},
\label{eq:recursive}
\end{equation}
where $N(t)=7t+1\approx 7t$ (when $t\gg 0$).

Since $n_0(t)=0$ when $t>0$, the time evolution of $n_1(t)$ is approximately written as
\begin{equation}
\frac{\mathrm{d}}{\mathrm{d} t}n_1(t) = -\frac{n_1(t)}{7t}.
\label{eq:n_1(t)}
\end{equation}
The solution of Eq. (\ref{eq:n_1(t)}) with the initial condition of $n_1(1)=8$ is
\begin{equation}
n_1(t)=\frac{8}{t^{1/7}}.
\label{eq:initial}
\end{equation}

We solve the recursive equation Eq. (\ref{eq:recursive}) with this initial condition Eq. (\ref{eq:initial}), and finally obtain
\begin{equation}
n_D(t)=\frac{8^D}{7^{D-1}t^{1/7}}\frac{(\ln t)^{D-1}}{(D-1)!}.
\label{eq:final}
\end{equation}

Therefore, the surface area $A$ of the simple FCD model at time $T$ is described as
\begin{eqnarray}
\begin{split}
A & = \sum_{D=1}^{D_{\max}}6\left[ \left( \frac{1}{2} \right)^{D} L \right]^2 n_D(T) \\
& = 12 L^2 T^{1/7}\frac{\Gamma(D_{\max},\frac{2}{7}\ln T)}{\Gamma(D_{\max})},
\end{split}
\end{eqnarray}
where $D_{\max}$ denotes the maximum number of cube division (i.e., the minimum cell length is $(1/2)^{D_{\max}}$).
$\Gamma(x)$ and $\Gamma(x,y)$ indicate the Euler's gamma function and incomplete gamma function, respectively.

Assuming $D_{\max} \gg 0$ (i.e., the minimum cell length $(1/2)^{D_{\max}} \ll L$), the above equation is approximately written as
\begin{equation}
A = 12 L^2 T^{1/7}.
\label{eq:surface_area}
\end{equation}

The relative frequency $P(D)$ (i.e., probability distribution) of cubes with length of $(1/2)^DL$ (at time $T$) is approximately equivalent to $n_D(T)/(7T)$; thus, we have
\begin{equation}
P(D) \simeq \frac{\lambda^{D-1}}{(D-1)!}\exp(-\lambda),
\label{eq:logpoisson}
\end{equation}
where $\lambda = (8/7)\ln T$.
Equation (\ref{eq:logpoisson}) indicates that $P(D)$ is a Poisson distribution with the mean $D$ of $\lambda + 1$, suggesting a geometric parameter $x$ of divided cubes (e.g., length $l$, surface area $a$, volume $v$) follows a log-Poisson distribution because $D \propto \ln x$.

\subsection{Numerical evaluations}
\label{sec:evaluation}

To evaluate the validity of these theoretical results, we performed numerical simulations.

We first compared the relationship between the surface area $A$ and volume $V$ (i.e., $=L^3$) between our theory and numerical simulation.
Supposing that $T=\lfloor V^{\xi} \rfloor$, we immediately obtain the relationship between $A$ and $V$ from Eq. (\ref{eq:surface_area}): 
\begin{equation}
A=12V^{\frac{14+3\xi}{21}}.
\label{eq:A-V_relationship}
\end{equation}

\begin{figure}[tbhp]
	\begin{center}
	\includegraphics{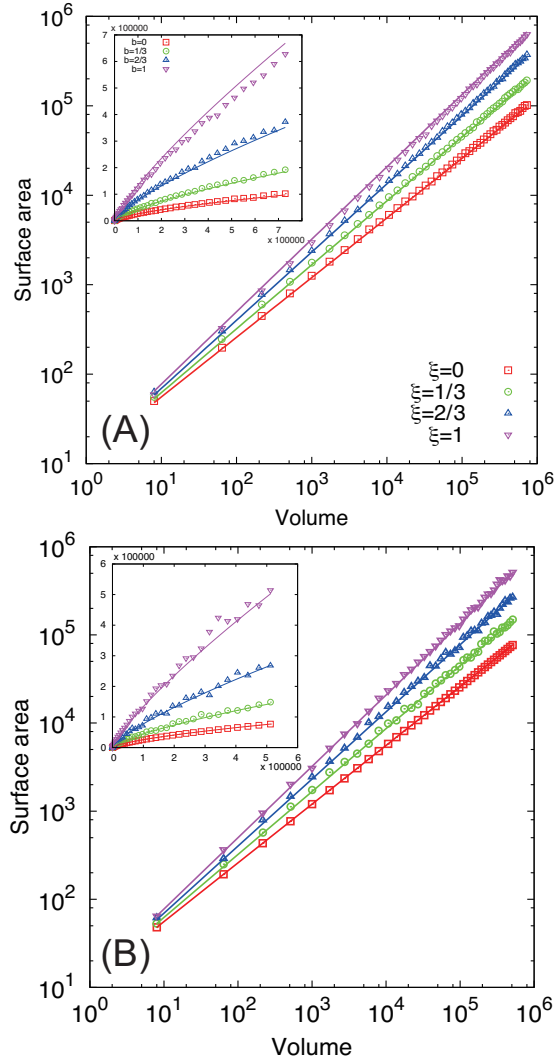}
	\caption{(Color online)
	Relationships between surface area and volume of the simple FCD model (A) and general FCD model (B), in which the parameter $p$ is drawn from the uniform distribution $U(0,1)$.
	These simulations consider that $T=\lfloor V^{\xi} \rfloor$.
	Symbols denote numerical results, which are averaged over 100 realizations.
	Lines are drawn from Eq. (\ref{eq:A-V_relationship}).
	Insets represent the same plots in normal scale. 
	}
	\label{fig:mass_area}
	\end{center}
\end{figure}

The prediction from Eq. (\ref{eq:A-V_relationship}) is in excellent agreement with the numerical results in both cases of the simple and general FCD models (Fig. \ref{fig:mass_area}).
In this case, the parameter $p$ in the general FCD model was drawn from the uniform distribution $U(0,1)$ when dividing a cuboid.

We also confirmed that Eq. (\ref{eq:A-V_relationship}) can estimate the $A$--$V$ relationships numerically obtained from the general FCD model, in which the parameter $p$ is sampled from the beta distribution Beta$(2,2)$, a bell-shaped distribution whose random variables range between 0 and 1, when separating a cuboid.
The prediction accuracy in this case is almost equal to that in the case of the uniform distribution.

These results indicate that the analytical solutions of the simple FCD model can be applicable to the general FCD model when the parameter $p$ is a random variable drawn from a probability distribution with the mean of 0.5 and limited variance.
The applicability of the analytical result obtained from the simple model to the general model is due to equivalents of the central limit theory.

We next compared the predicted relative frequency (probability density) of geometric parameters with numerical simulations.
As an example of geometric parameters, we focused on cube volumes $v$.
Using the relational expression $D=-\ln(v/V)/\ln 8$, the relative frequency can be estimated from Eq. (\ref{eq:logpoisson}).

\begin{figure}[tbhp]
	\begin{center}
	\includegraphics{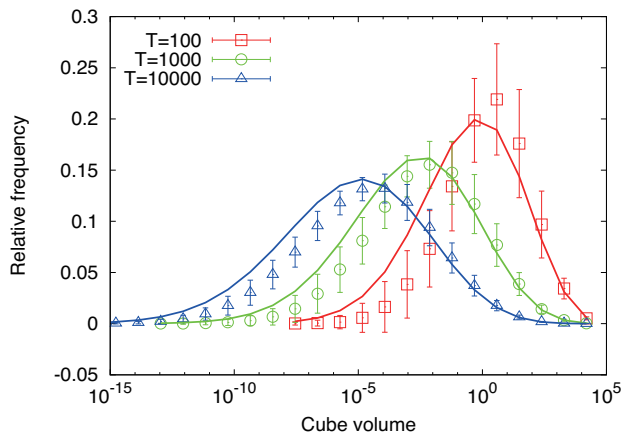}
	\caption{(Color online) Relative frequencies of cube volumes $v$ in the simple FCD model with $L=40$ and $T=100$, $1000$, and $10000$.
	Symbols denote numerical results, which are averaged over 100 realizations.
	Error bars indicate the standard deviations.
	Lines are drawn from Eq. (\ref{eq:logpoisson}) using $D=-\ln(v/V)/\ln 8$.
	}
	\label{fig:cell_volume_distribution}
	\end{center}
\end{figure}

As shown in Fig. \ref{fig:cell_volume_distribution}, our theoretical predictions are in good agreement with numerical results although the prediction accuracy becomes lower for abundant cubes with a smaller volume due to the several approximations (see Sec. \ref{sec:analysis}).
However, this limitation poses little problems when estimating the total surface area $A$ of the model because the theoretical results are in excellent agreement with numerical results (Fig. \ref{fig:mass_area}). 

We also confirmed that Eq. (\ref{eq:logpoisson}) can predict numerical results in both cases of cube lengths $l$ and cube surface areas $a$.
Substituting the relational expression $D=-\ln(l/L)/\ln 2$ ($D=-\ln(a/A_0)/\ln 4$, where $A_0=6L^2$) into Eq. (\ref{eq:logpoisson}), we can obtain the theoretical distributions of $l$ ($a$).

\subsection{Comparison with empirical data}
\label{sec:comparison}
Assuming $B\propto A$ and $M \propto V$ \citep{Kozlowski2003,White2003}, we can discuss the origin of the allometric scaling of metabolic rate using the FCD model; however, the parameter $T$ needs to be estimated by considering some constraints.
We here estimate $T$ using empirical results, and derive the allometric scaling laws of metabolic rate.

For example, two extremely possible strategies \citep{Savage2007} may be useful: (i) average cell (cube or cuboid) mass $\langle m \rangle$ remains constant while average cellular metabolic rate $\langle b \rangle$ changes, or (ii) $\langle b \rangle$ remains fixed while $\langle m \rangle$ varies, where $\langle x \rangle$ indicates the mean of $x$.
According to a previous study \citep{Savage2007}, we assume that $\langle m \rangle \propto \langle v \rangle$ and $\langle b \rangle \propto \langle a \rangle$, where $\langle v \rangle$ and $\langle a \rangle$ correspond to average cellular volume and average cellular surface area $\langle a \rangle$, respectively.
In this case, the parameter $T$ can be estimated as follows:
In particular, the strategy (i) leads to $T\propto M$ because $\langle m \rangle \propto V/T$.
Moreover, the strategy (ii) derives $T\propto M^{7/9}$ because $\langle a \rangle = 12M^{2/3}T^{1/7}/(7T)$.
Therefore, we find
\begin{equation}
B \propto \left\{ \begin{array}{ll}
M^{17/21}=M^{0.81} & (\mathrm{strategy \ (i)}) \\
M^{7/9}=M^{0.78} & (\mathrm{strategy \ (ii)})
\end{array}
\right.
\end{equation}

On the other hand, if $T$ is proportional to the number of cells $N_c$ because of the definition, our model predicts $N_c\propto M$ and $N_c\propto M^{0.78}$ in the case of the strategies (i) and (ii), respectively.
Theoretical predictions are in agreement with the observed results (see Table 1 in \citep{Savage2007}). 

However, since the distributions of $a$ and $v$ are log-normal like (Fig. \ref{fig:cell_volume_distribution} and Eq. (\ref{eq:logpoisson})), it remains possible that the estimation of $T$ using means is unsuitable.
When considering the multiplicative process such as the FCD model (cell divisions), from a statistical perspective, geometric means should also be considered.
Thus, we further considered 2 strategies by modifying the strategies (i) and (ii): (iii) logarithmic mean of cell mass $\langle \ln m \rangle$ remains constant while logarithmic mean of cellular metabolic rate $\langle \ln b \rangle$ changes, or (iv) $\langle \ln b \rangle$ remains fixed while $\langle \ln m \rangle$ varies.

$\langle \ln v \rangle$ and $\langle \ln a \rangle$ in the simple FCD model are obtained as
\begin{eqnarray}
\begin{split}
\langle \ln v \rangle & = \sum_{D=1}^{\infty} \ln \left[ \left\{ \left( \frac{1}{2}\right)^d L \right\}^3 \right]P(D)\\
& = 3\left(\ln\frac{L}{2}-\frac{8}{7}\ln T \ln 2\right)
\end{split}
\label{eq:ln_m}
\end{eqnarray}
and
\begin{eqnarray}
\begin{split}
\langle \ln a \rangle & = \sum_{D=1}^{\infty} \ln \left[ 6\left\{ \left( \frac{1}{2}\right)^d L \right\}^2 \right]P(D)\\
& = 2\left(\ln\frac{L}{2}-\frac{8}{7}\ln T \ln 2 + 3\right),
\end{split}
\label{eq:ln_a}
\end{eqnarray}
respectively.

Supposing that $\langle \ln b \rangle \propto \langle \ln a \rangle$ and $\langle \ln m \rangle \propto \langle \ln v \rangle$, both strategies (iii) and (iv) lead to $T\propto M^{7/(24\ln2)}$.
Thus, we obtain
\begin{equation}
B \propto M^{\frac{2}{3}+\frac{1}{24\ln2}} = M^{0.73} \mathrm{ \ (strategies \ (iii) \ \& \ (iv))}.
\end{equation}

The scaling exponent ranges from 0.73 to 0.81 according to the biological strategies, and it may fall within 95\% confidential intervals of scaling exponents estimated from empirical data (e.g., see \citep{Savage2004}).
In addition to this, this result is consistent with a conclusion, derived by \citet{Kolokotrones2010}, that the scaling exponent is around 0.8 rather than $3/4$ when only focusing on mammals with higher $M$.
This finding may not contradict the WEB model because this model is known to show the scaling exponent of 0.81 (not 3/4) when considering finite-size corrections \citep{Savage2008}.

The relationship between the C-value (i.e., genome size) and $M$ \citep{Kozlowski2003} is also useful for estimating the parameter $T$; in particular, the C-value has been known to show an approximately linear relationship with cell size \citep{Gregory2001}.
This previous study has reported that C-value $\propto M^{\beta}$ with $\beta \in (-0.05,0.06)$; thus, it encourages an extension of the strategies (i) and (iii).
In particular, we considered the following extended strategies: (I) $\langle m \rangle \propto M^{\beta}$ and (III) $\langle \ln m \rangle \propto \beta \ln M$.
When $\beta = 0$, the strategies (I) and (III) are equivalent to the strategies (i) and (iii), respectively.

Using the relational expression $\langle m \rangle \propto V/T$ and Eq. (\ref{eq:ln_m}), we can obtain the allometric relationships of $B$ based on the strategies (I) and (III), respectively.
In particular, the scaling exponents are described as
\begin{equation}
\gamma = \frac{17-3\beta}{21} \mathrm{\ (strategy \ (I))}
\label{eq:FCD_I}
\end{equation}
and
\begin{equation}
\gamma = \frac{2}{3}+\frac{1-\beta}{24\ln 2} \mathrm{\ (strategy \ (III))},
\label{eq:FCD_III}
\end{equation}
respectively.

\begin{figure}[tbhp]
	\begin{center}
	\includegraphics{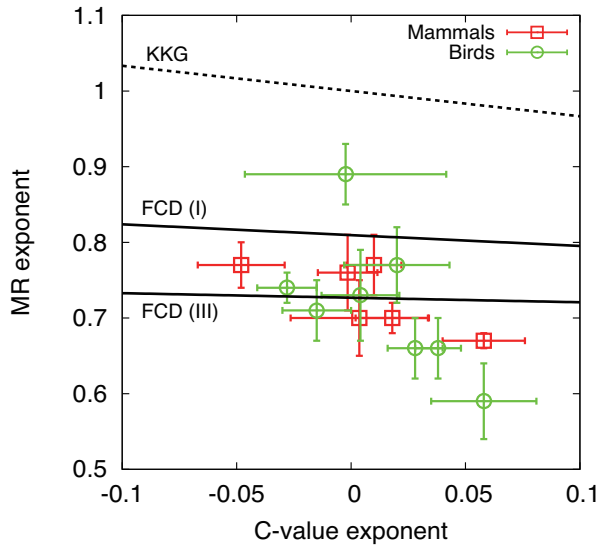}
	\caption{(Color online) Relationship between metabolic rate (MR) exponent $\gamma$ and C-value exponent $\beta$.
	Symbols denote the empirical data, obtained from the study by \citet{Kozlowski2003}, and they indicate lineages (orders).
	Error bars indicate standard errors.
	The solid lines are drawn from Eqs. (\ref{eq:FCD_I}) and (\ref{eq:FCD_III}) (our (FCD) model), respectively.
	The dashed line is obtained from Eq. (\ref{eq:KKG_model}) (KKG model).
	}
	\label{fig:comparison_exponent}
	\end{center}
\end{figure}

We evaluated these theoretical predictions with the empirical data \citep{Kozlowski2003} (Fig. \ref{fig:comparison_exponent}).
For comparison, in addition, we also investigated the $\gamma$--$\beta$ relationship predicted from the KKG model.
The total surface area of the KKG model is described as $A\propto \langle m \rangle^{2/3}N_c$, where $N_c\propto M/\langle m \rangle$.
Since $\ln \langle m \rangle = \langle \ln m \rangle \propto \beta \ln m$ because all cell sizes are identical (i.e., the variance of cell size is zero) in the KKG model, the $\gamma$--$\beta$ relationship, derived according to the strategies (I) and (III), is the same as follows:
\begin{equation}
\gamma=1-\frac{\beta}{3} \mathrm{\ (KKG \ model)}.
\label{eq:KKG_model}
\end{equation}

Our theoretical predictions are in good agreement with the empirical data, and they show higher prediction accuracy, compared with the KKG model.
Our theory can also explain a negative correlation between metabolic rate exponent $\gamma$ and C-value exponent $\beta$.
However, the theoretical prediction of this tendency may be qualitative in this case.
In particular, the FCD model could not quantitatively predict the $\gamma$--$\beta$ relationships for a larger $\beta$.
This may be because the exponent $\beta$ was estimated from average cell size (i.e., $\langle m \rangle$) calculated using a few number of cells.
That is, it remains possible that the assumption that $\langle m \rangle \propto M^{\beta}$ and $\langle \ln m \rangle \propto \beta \ln M$ is not satisfied.
A more careful examination of the theoretical predictions may require an estimation of $\beta$ using a larger number of cells in each organism.

We can also consider an extension of the strategies (ii) and (iv) according to the $M$-dependence of cell size \citep{Kozlowski2003}: (II) $\langle a \rangle \propto M^{\beta}$ and (IV) $\langle \ln a \rangle \propto \beta \ln M$.
Note that $\beta$ does not correspond to the C-value exponent in this case.  
When $\beta = 0$, the strategies (II) and (IV) are equivalent to the strategies (ii) and (iv), respectively.

According to the relational expression $\langle a \rangle = 12M^{2/3}T^{1/7}/(7T)$ and Eq. (\ref{eq:ln_a}), we find the allometric relationships of $B$ based on the strategies (II) and (IV), respectively.
Especially, the scaling exponents are obtained as
\begin{equation}
\gamma = \frac{14-3\beta}{18} \mathrm{\ (strategy \ (II))}
\end{equation}
and
\begin{equation}
\gamma = \frac{2}{3}+\frac{2-3\beta}{48\ln 2} \mathrm{\ (strategy \ (IV))},
\end{equation}
respectively.

\subsection{Ubiquity of the 3/4-power law}
\label{sec:ubiquity}
The strategies (I)--(IV) may be mixed among real-world living organisms.
Furthermore, a huge variety of $\beta$ may be observed.
What kind of distributions does the metabolic rate exponent $\gamma$ follow in this situation?

To consider various possible situations, we assumed the following condition: the strategies (I)--(IV) are uniformly observed, and $\beta$ is drawn from the uniform distribution $U(-1,1)$.

\begin{figure}[tbhp]
	\begin{center}
	\includegraphics{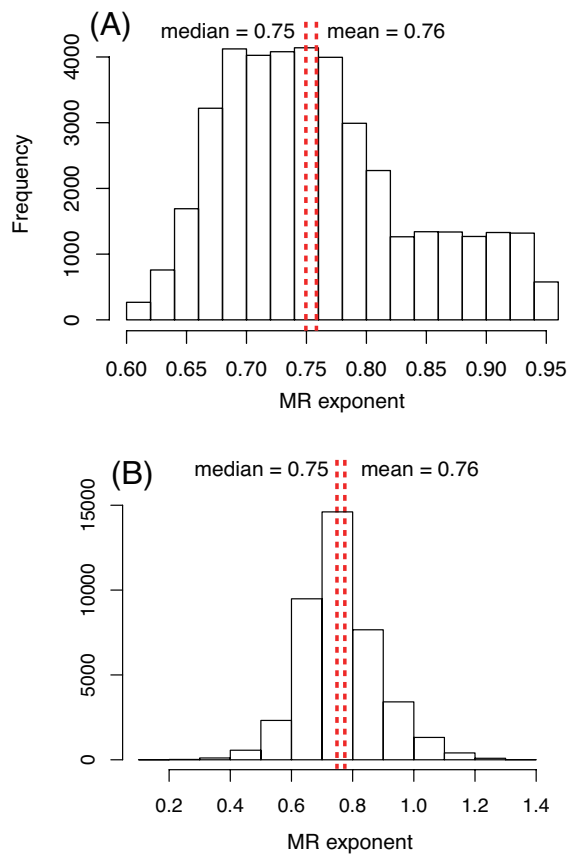}
	\caption{(Color online) Distribution of metabolic rate (MR) exponent $\gamma$ obtained from the FCD model according the strategies (I)--(IV), in which $\beta$ is drawn from the uniform distribution $U(-1,1)$ (A) and the standard normal distribution $N(0,1)$ (B).
	}
	\label{fig:exponent_distribution}
	\end{center}
\end{figure}

The 40000 trials performed under this condition showed a bell-shape distribution of $\gamma$ with the mean of 0.76 (the median of 0.75) (Fig. \ref{fig:exponent_distribution} A) although the strategies and $\beta$ are uniformly selected.

In terms of evolutionary dynamics, it may be more realistic to consider that $\beta$ follows a Gaussian-like distribution.
We also performed a similar analysis by assuming that $\beta$ is sampled from a normal distribution with the mean of 0 and variance of 1 (i.e., $N(0,1)$), and again found that the metabolic rate exponent follows a bell-shape distribution of $\gamma$ with the mean of 0.76 (the median of 0.75) (Fig. \ref{fig:exponent_distribution} B). 

This result implies that the 3/4-power law can be emerged from a number of different biological strategies.
Although a previous geometric perspective \citep{Okie2013,Hirst2014} shows that the metabolic rate exponent $\gamma$ can be variable, it cannot answer why the 3/4-power law is universally found in nature \citep{Savage2004}.
The FCD model may answers this question.

\section{Conclusion}
We proposed a simple geometry model for the emergence of allometric scaling of metabolic rate.
Several analytical solutions of this model of geometric parameters were provided (see Sec. \ref{sec:analysis}), and their validity was confirmed by numerical simulations (see Sec. \ref{sec:evaluation}). 

Using this model, we discussed a possible origin of allometric scaling of metabolic rate, and concluded that the approximately 3/4-power law can be also derived from the context of total surface area of cell sizes by considering fractal-like (or hierarchical) organization of living organisms, contrary to criticisms (e.g., \citep{Brown2005}).
The theoretical predictions obtained from this model are in good agreement with empirical results (see Sec. \ref{sec:comparison}).
Unlike previous models, in particular, this model can explain the important observations: the metabolic rate exponent $\gamma$ shows a central tendency of 3/4 (see Sec. \ref{sec:ubiquity}) although it can be variable \citep{Okie2013,Hirst2014}.
These results implies that a hypothesis total surface area of cells determines metabolic rate evidently shows promise for explaining the origin of the allometric scaling.

The FCD model was proposed inspired by the KKG model \citep{Kozlowski2003} and WBE model \citep{West1997}; in particular, it considers the effect of cell size (KKG model) on metabolic rate and the concept of fractal-like organization (WBE model) (see Sec. \ref{sec:model} for details).
Although \citet{Kozlowski2004} and \citet{Brown2005} debated over the validity of each other's models, the FDC model suggests that these models are complementary rather than conflicting.
In particular, the FCD model shows a fractal-like organization (heterogeneous distribution) of cells is required that when deriving the approximately 3/4-power law in the context of geometry (the KKG model). 

The FCD model implies the importance of heterogeneous distributions (e.g., log-normal distributions of cell sizes for the allometric scaling.
In this paper, we could not test this hypothesis because cell size distributions are less understood.
However, log-normal distributions are widely observed in nature \citep{Limpert2001}, and several previous studies suggest that the size of red cell, which is related to metabolic rate, approximately follows a log-normal distribution \citep{Bell1985,McLaren1991}.
Therefore, we expect that this hypothesis provides a new perspective for the origin of metabolic scaling.
Although several studies \citep{Kozlowski2003,Starostova2009} have reported the importance of cell size for scaling laws of metabolic rate, there is little study focusing on such size distributions.
Our study encourages an evaluation of the relationship between the scaling laws of metabolic rate and cell size distributions (i.e., single-cell measurements).

However, more careful examinations may be required.
The relationship between the surface area and metabolic rate is still unclear because of several reasons.
For example, a number of factors influence metabolic rates.
The rate of supply and transportation of resources within the body is considered as the fundamental constraint on metabolic rate; however, the FCD model does not consider this constraint.
Although the total surface area of mitochondrial membranes on which the majority of ATP synthesis occurs may be correlated with metabolic rate, it is not conformed whether such a correlation is also observed in the other types of cells.

Even though limitations are inherent in our model due to the simplicity of the model, we believe that our model still serves to provide intuitive and unique insights into the origin of allometric scaling in metabolic rate.

\section*{Acknowledgments}
This study was supported by a Grant-in-Aid for Young Scientists (A) from the Japan Society for the Promotion of Science (no. 25700030).
K.T. was partly supported by Chinese Academy of Sciences Fellowships for Young International Scientists (no. 2012Y1SB0014), and the International Young Scientists Program of the National Natural Science Foundation of China (no. 11250110508).
K.T. thanks Dr. Kohei Koyama for providing useful information, and he is also obliged to anonymous reviewers for helpful comments and suggestions.


\begin{thebibliography}{00}

\bibitem[Banavar et al., 1999]{Banavar1999}
Banavar, J.R., Maritan, A., Rinaldo, A., 1999. Size and form in efficient transportation networks. Nature 399, 130--132. doi:10.1038/20144  

\bibitem[Banavar et al., 2010]{Banavar2010}
Banavar, J.R., Moses, M.E., Brown, J.H., Damuth, J., Rinaldo, A., Sibly, R.M., Maritan, A., 2010. A general basis for quarter-power scaling in animals. Proc. Natl. Acad. Sci. U. S. A. 107, 15816--15820. doi:10.1073/pnas.1009974107

\bibitem[Banavar et al., 2014]{Banavar2014}
Banavar, J.R., Cooke, T.J., Rinaldo, A., Maritan, A., 2010. Form, function, and evolution of living organisms. Proc. Natl. Acad. Sci. U. S. A. 111, 3332--3337. doi:10.1073/pnas.1401336111

\bibitem[Bell, 1985]{Bell1985}
Bell, G.M., Fowler, J.S., 1985. Red cell population distributions in healthy dogs. Res. Vet. Sci. 38, 220--225.  

\bibitem[Brown et al., 2004]{Brown2004}
Brown, J.H., Gillooly, J.F., Allen, A.P., Savage, V.M., West, G.B., 2004. Toward a metabolic theory of ecology. Ecology 85, 1771--1789.

\bibitem[Brown et al., 2005]{Brown2005}
Brown, J.H., West, G.B., Enquist, B.J., 2005. Yes , West , Brown and Enquist's model of allometric scaling is both mathematically correct and biologically, Funct. Ecol. 19, 735--738.  


\bibitem[Darveau et al., 2002]{Darveau2002}
Darveau, C.-A., Suarez, R.K., Andrews, R.D, Hochachka, P.W., 2002. Allometric cascade as a unifying principle of body mass effects on metabolism. Nature 417, 166--170. doi:10.1038/417166a  

\bibitem[Demetrius and Tuszynski, 2010]{Demetrius2010}
Demetrius, L., Tuszynski, J.A., 2010. Quantum metabolism explains the allometric scaling of metabolic rates. J. R. Soc. Interface 7, 507--514. doi:10.1098/rsif.2009.0310  

\bibitem[Enquist et al., 2007]{Enquist2007}
Enquist, B.J., et al., 2007. Biological scaling: does the exception prove the rule? Nature 445, E9--E10. doi:10.1038/nature05548  

\bibitem[Gregory, 2001]{Gregory2001}
Gregory, T.R., 2001. Coincidence, coevolution, or causation? DNA content, cell size, and the C-value enigma, Biol. Rev. Camb. Philos. Soc. 76, 65--101.  

\bibitem[Hirst et al., 2014]{Hirst2014}
Hirst, A.G., Glazier, D.S., Atkinson, D., 2014. Body shape shifting during growth permits tests that distinguish between competing geometric theories of metabolic scaling. Ecol. Lett. 17, 1274--1281.  

\bibitem[Jetz et al., 2004]{Jetz2004}
Jetz, W., Carbone, C., Fulford, J., Brown, J.H., 2004. The scaling of animal space use. Science 306, 266--268. doi:10.1126/science.1102138  

\bibitem[Kobayashi et al., 2011]{Kobayashi2011}
Kobayashi, N., Kuninaka, H., Wakita, J., Matsushita, M., 2011. Statistical features of complex systems: toward establishing sociological physics. J. Phys. Soc. Japan 80, 072001. doi:10.1143/JPSJ.80.072001  

\bibitem[Kolokotrones et al., 2010]{Kolokotrones2010}
Kolokotrones, T., Savage, V., Deeds, E.J., Fontana, W., 2010. Curvature in metabolic scaling. Nature 464, 53--756. doi:10.1038/nature08920  

\bibitem[Koz{\l}owski and Konarzewski, 2004]{Kozlowski2004}
Koz{\l}owski, J., Konarzewski, M., 2004. Is West, Brown and Enquist's model of allometric scaling mathematically correct and biologically relevant? Funct. Ecol. 18, 283--289. doi:10.1111/j.0269-8463.2004.00830.x  

\bibitem[Koz{\l}owski et al., 2003]{Kozlowski2003}
Koz{\l}owski, J., Konarzewski, M., Gawelczyk, A.T., 2003 Cell size as a link between noncoding DNA and metabolic rate scaling. Proc. Natl. Acad. Sci. U. S. A. 100, 14080--14085. doi:10.1073/pnas.2334605100  

\bibitem[Limpert et al., 2001]{Limpert2001}
Limpert, E., Stahel, W.A., Abbt, M., 2001. Log-normal sistributions across the sciences: keys and clues. Bioscience 51, 341--352.

\bibitem[Makarieva et al., 2008]{Makarieva2008}
Makarieva, A.M., et al., 2008. Mean mass-specific metabolic rates are strikingly similar across life's major domains: Evidence for life's metabolic optimum. Proc. Natl. Acad. Sci. U. S. A. 105, 16994--16999. doi:10.1073/pnas.0802148105  

\bibitem[McLaren et al., 1991]{McLaren1991}
McLaren, C.E., Wagstaff, M., Brittenham, G.M., Jacobs, A., 1991. Detection of two-component mixtures of lognormal distributions in grouped, doubly truncated data: analysis of red blood cell volume distributions. Biometrics 47, 607--622.  

\bibitem[Okie, 2013]{Okie2013}
Okie, J., 2013. General models for the spectra of surface area scaling strategies of cells and organisms: fractality, geometric dissimilitude, and internalization. Amer. Nat. 181, 421--439. doi:10.1086/669150  

\bibitem[Porter and Brand, 1993]{Porter1993}
Porter, R.K., Brand, M.D., 1993. Body mass dependence of H$^{+}$ leak in mitochondria and its relevance to metabolic rate. Nature 362, 628--630. doi:10.1038/362628a0  

\bibitem[Price et al., 2007]{Price2007}
Price, C.A., Enquist, J.B., Savage, V.M., 2007. A general model for allometric covariation in botanical form and function. Proc. Natl. Acad. Sci. U. S. A. 104, 13204--13209. doi:10.1073/pnas.0702242104  

\bibitem[Price et al., 2012]{Price2012}
Price, C.A., et al., 2012. Testing the metabolic theory of ecology. Ecol. Lett. 15, 1465--1474. doi:10.1111/j.1461-0248.2012.01860.x  

\bibitem[Reich et al., 2006]{Reich2006}
Reich, P.B., Tjoelker, M.G., Machado, J.-L., Oleksyn, J., 2006. Universal scaling of respiratory metabolism, size and nitrogen in plants. Nature 439, 457--61. doi:10.1038/nature04282  

\bibitem[Savage et al., 2008]{Savage2008}
Savage, V.M., Deeds, E.J., Fontana, W., 2008. Sizing up allometric scaling theory. PLoS Comput. Biol. 4, e1000171. doi:10.1371/journal.pcbi.1000171  

\bibitem[Savage et al., 2004]{Savage2004}
Savage, V.M., et al., 2004. The predominance of quarter-power scaling in biology. Funct. Ecol. 18, 257--282. doi:10.1111/j.0269-8463.2004.00856.x  

\bibitem[Savage et al., 2007]{Savage2007}
Savage, V.M., et al., 2007. Scaling of number, size, and metabolic rate of cells with body size in mammals, Proc. Natl. Acad. Sci. U. S. A. 104, 4718--4723. doi:10.1073/pnas.0611235104  

\bibitem[Speakman, 2005]{Spearkman2005}
Speakman, J.R., 2005. Body size, energy metabolism and lifespan. J. Exp. Biol. 208, 1717--1730. doi:10.1242/jeb.01556  

\bibitem[Starostov\'a et al., 2009]{Starostova2009}
Starostov\'a, Z., Kubicka, L., Konarzewski, M., Koz{\l}owski, J., Kratochv\'il, L., 2009. Cell size but not genome size affects scaling of metabolic rate in eyelid geckos. Am. Nat. 174, E100--E105. doi:10.1086/603610  

\bibitem[Szarski, 1983]{Szarski1983}
Szarski, H., 1983, Cell size and the concept of wasteful and frugal evolutionary strategies, J. Theor. Biol. 105, 201--209. doi:10.1016/S0022-5193(83)80002-2.  

\bibitem[Takemoto, 2012]{Takemoto2012}
Takemoto, K., 2012. Current understanding of the formation and adaptation of metabolic systems based on network theory. Metabolites 2, 429--457. doi:10.3390/metabo2030429  

\bibitem[Takemoto and Oosawa, 2012]{Takemoto2012b}
Takemoto, K., Oosawa, C., 2012. Modeling for evolving biological networks. in: Statistical and Machine Learning Approaches for Network Analysis, pp. 77--108. doi:10.1002/9781118346990.ch3  

\bibitem[West et al., 1997]{West1997}
West, G.B., Brown, J.H., Enquist, B.J., 1997. A general model for the origin of allometric scaling laws in biology. Science 276, 122--126.  

\bibitem[West et al., 1999]{West1999}
West, G.B., Brown, J.H., Enquist, B.J., 1999. The fourth dimension of life: fractal geometry and allometric scaling of organisms. Science 284, 1677--1679.  

\bibitem[West et al., 2002]{West2002}
West, G.B., Woodruff, W.H., Brown, J.H., 2002. Allometric scaling of metabolic rate from molecules and mitochondria to cells and mammals. Proc. Natl. Acad. Sci. U. S. A. 99, 2473--2478. doi:10.1073/pnas.012579799  

\bibitem[White and Seymour, 2003]{White2003}
White, C.R., Seymour, R.S., 2003. Mammalian basal metabolic rate is proportional to body mass$^{2/3}$. Proc. Natl. Acad. Sci. U. S. A. 100, 4046--4049. doi:10.1073/pnas.0436428100  

\bibitem[Yamamoto and Yamazaki, 2013]{Yamamoto2013}
Yamamoto, K., Yamazaki, Y., 2013. Group-separation effect in cell-size distribution of origami crease patterns. J. Phys. Soc. Japan 82, 044803. doi:10.7566/JPSJ.82.044803 

\end{thebibliography}
\end{document}